\documentclass[11pt,a4paper]{article}
\usepackage[hyperref]{acl2020}
\usepackage[symbol]{footmisc}
\renewcommand{\thefootnote}{\fnsymbol{footnote}}

\usepackage{graphicx}
\usepackage[T1]{fontenc}

\usepackage{tikz}
\usetikzlibrary{positioning}
\usepackage{times}
\usepackage{latexsym}

\usepackage{microtype}

\aclfinalcopy 

\title{End-to-End Speech Recognition and Disfluency Removal with Acoustic Language Model Pretraining}
\author{
        Saksham Bassi\footnote[2]{All authors contributed equally.} \\ sb7787 \And
        Giulio Duregon\footnote[2]{All authors contributed equally.} \\ gjd9961 \And
        Siddhartha Jalagam\footnote[2]{All authors contributed equally.} \\ scj9994 \And
        David Roth\footnote[2]{All authors contributed equally.} \\ dsr331
    }

\begin{document}
\maketitle

\begin{abstract}
The SOTA in transcription of disfluent and conversational speech has in recent years favored two-stage models, with separate transcription and cleaning stages. We believe that previous attempts at end-to-end disfluency removal have fallen short because of the representational advantage that large-scale language model pretraining has given to lexical models. Until recently, the high dimensionality and limited availability of large audio datasets inhibited the development of large-scale self-supervised pretraining objectives for learning effective audio representations, giving a relative advantage to the two-stage approach, which utilises pretrained representations for lexical tokens. In light of recent successes in large scale audio pretraining, we revisit the performance comparison between two-stage and end-to-end model and find that audio based language models pretrained using weak self-supervised objectives match or exceed the performance of similarly trained two-stage models, and further, that the choice of pretraining objective substantially effects a model's ability to be adapted to the disfluency removal task.\footnote[1]{Code: \url{https://github.com/davidsroth/hubert-disfl/}}
\end{abstract}

\let\thefootnote\relax\footnotetext{$^\dagger$ All authors contributed equally.}

\section{Introduction}

Conversations, dialogue, and spontaneous speech differ from text sources in that they often contain errors that are self-corrected throughout a given utterance. 
Producing clean transcriptions of these signals is often difficult, requiring the model to identify which segments to include and omit. Popular modern approaches have addressed this problem using a two-stage transcription process- first, the sequence is transcribed verbatim to a sequence of text tokens, which is then fed to a separately trained text model to remove disfluent or self-corrected sections of speech \cite{jamshid-lou-etal-2019-neural} In this two stage formulation, a disfluency model is learned on text tokens alone-- audio features from the original signal are used only for producing text tokens during the first stage and not included during disfluency removal. In doing so, the fluency model cannot access intonation, tempo and prosody cues from the original audio signal which can be informative for the disfluency removal task. 

\begin{figure}[t]
    \centering
    \scalebox{0.65}{
         \begin{tikzpicture}[node distance=7em, every node/.style = {shape=rectangle, draw,
                             minimum height=1.5em, line width=1pt,
                             align=center, text height=3mm}, every edge/.append style = {line width=1pt}]
             \node[text width=85mm, fill=yellow!40] (f1) { \textcolor{gray}{"I mean {\color{red} it was just} it was probably one of the most strengthening things"}};
             
             \node[below=1.5em of f1, text width=45mm, fill=blue!30] (c1) {Parser};
             
             \node[below=1.5em of c1, text width=75mm, fill=yellow!40] (d1) {\textcolor{gray}{"I mean it was probably one of the most strengthening things"}};
             
             \node[below=1.5em of d1, xshift=2em, text width=15mm, fill=orange!40] (d2) {Audio};
             
             
             \node[below=5em of d1, text width=35mm, text height=1em, xshift=0.0em, fill=blue!30] (e1) {\textbf{HuBERT model + CTC Decoder}};
             
             \node[below=2.5em of e1, text width=45mm, fill=green!40] (e2) {Fluent transcripts};
             
             \node[below=1em of d1, node distance=2.6em, xshift=-10.5em, yshift=0em, draw=none] {\scalebox{.7}{Target}};
             
             \node[below=1em of d2, node distance=2.6em, xshift=-3.0em, yshift=1.5em, draw=none] {\scalebox{.7}{Input}};
             
             \draw [->] (f1) edge (c1);
             \draw [->] (c1) edge (d1);
             \draw [->] (e1) edge (e2);

            \draw [->] (e1) -- coordinate[pos=0.5] (mid) (e2);
            \draw [->, bend right=30] (d2) edge (e1);
            \draw [->, bend right=80] (d1) edge (mid);
         \end{tikzpicture}
    }
    \caption{Model Flow of the end-to-end system}
    \label{fig:model_flow}
\end{figure}
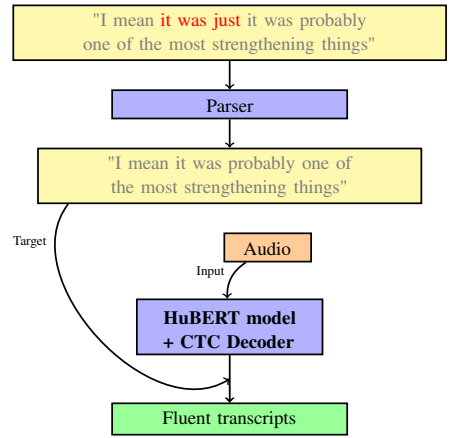
Many modern applications, including Amazon's Alexa, Apple's Siri and Google's Voice Search use speech recognition software to convert speech signals into a text format that is interpretable by an inference model to determine user intent. Spoken word audio is recorded as input, transformed into text representations via a transcription model, and mapped to outputs based on the interpreted meaning. These models often use text pretraining to improve their performance, but the divergence between written and spoken sources creates difficulties for interpreting disfluent speech, making proper handling of speech disfluencies difficult for downstream applications.

To bridge the gap between the written texts that inference models are trained on and the spontaneous speech captured from downstream users, modern ASR systems can include a text-based disfluency detection step, in which tokenized lexical representations produced by a transcription model are fed to a disfluency classification step to remove disfluencies before ingested by an inference model \cite{jamshid-lou-etal-2019-neural}. We believe that end-to-end approaches can utilize prosody cues for disfluency detection that are not captured by tokenized lexical representations and hope to improve their performance and/or identify scenarios where prosody information leads to better performance. 

We investigate an end-to-end approach, using a pretrained acoustic model to directly predict fluent transcripts of disfluent speech. We train our model on Switchboard \cite{switchboard} and evaluate using Word Error Rate (WER) and Character Error Rate (CER) on disfluency filtered text. We also investigate the effect that the choice of pretraining objective plays in our model's performance by comparing the fine-tuning performance of two Conformer \cite{Conformer} based models, one utilizing a constrastive Masked Language Modeling pretraining objective and one utilizing a lower dimensional clustering based objective.

\section{Background}

\subsection{Automatic Speech Recognition (ASR)}
Computational approaches to Automatic Speech Recognition go back as far as the 1950's. The earliest systems used template matching against known speech patterns to map sections of spoken audio to transcriptions, but were highly limited in their scope, only transcribing digits from a single speaker \cite{li2019vocal}. Later, more advanced techniques utilizing Hidden Markov Models (HMM) were used to map audio data into sequences of phonemes, which were translated into likely sequences of words using dynamic programming via beam search \cite{historyofasr}. Following the successes of Deep Learning approaches to image recognition and natural language processing in the 21st century, HMM systems evolved into Recurrent Neural Network (RNN) and Convolutional Neural Network approaches, which achieved impressive gains over their predecessors. 

Recent successes in image and language tasks with the Transformer architecture prompted exploration into self-attentive approaches to ASR, but the comparatively large sequence lengths of audio signals and the quadratic complexity inherent to Transformer models presented significant computational challenges for self-attentive audio models. In the last few years, models that use Convolutional Neural Networks (CNN) to downsample audio sequences before processing with transformers, combined with systems with larger computational capacity, enabled self-attentive approaches to audio, which have since overtaken RNNs and LSTMs as the strongest performers in many audio-based tasks (\citet{Conformer} ; \citet{Han2020ContextNet}). This approach has proved quite effective, with \textit{Conformer} achieving a state of the art performance on the LibriSpeech benchmark, achieving a WER of $1.9\%/3.9\%$ on the test\_clean / test\_other portions of the LibriSpeech test sets.

\subsection{Acoustic Model Pretraining}
Performance gains from self-supervised pretraining using Transformers have led to strong performance gains on many natural language tasks, prompting exploration into extensions to the pretraining framework for acoustic modeling. Here, as before, the high dimensionality of audio data presented difficulties in training these systems using a masked objective \cite{Chung2021w2vBERTCC}.
HuBERT \cite{HuBERT} is a pretraining framework that uses a Conformer \cite{Conformer} architecture to process sequences of audio frames and a kNN clustering step to provide a lower-dimensional, stable training signal for masked token prediction over high-dimensional audio features. The model is trained to predict a sequence of hidden states $[Z_1,Z_2,...,Z_t]$ over masked portions of the final Conformer output layer, where $Z_t$ is a C-class categorical variable corresponding to cluster assignments produced by an ensemble of clustering models that are iteratively refined during the training process. This pretraining formulation was found to produce speech token representations that significantly improve ASR performance, particularly in contexts with limited training data.

\subsection{Disfluency Detection}
Early attempts at using statistical modeling to identify disfluencies in spoken language used a combination of prosodic and lexical cues to detect errata and reparanda, but found greater gains from lexical representations (\citet{Baron02automaticpunctuation}, \citet{SnoverDS04}, \citet{shriberg1997prosody}).
The increasing availability of large, open text corpora fueled subsequent improvements to representations from deep architectures, making lexical representations increasingly the focal point for disfluency detection approaches (\citet{qian-liu-2013-disfluency}, \citet{wang-etal-2016-neural}, \citet{jamshid-lou-johnson-2020-improving}). Subsequent approaches to improve disfluency detection performance used data augmentation and semi-supervised objectives to expand the volume of "disfluent" lexical data available for training (\citet{Wang2020MultiTaskSL}, \citet{jamshid-lou-johnson-2020-improving}) and introduced secondary syntactic objectives for multi-task training (\citet{Lee2021AuxiliarySL}, \citet{honnibal-johnson-2014-joint}). \citeauthor{zayats} re-introduced prosody cues by providing prosody features learned through text-based distributional prediction alongside lexical features during training and inference. 


\subsection{End-to-End ASR and Disfluency Detection Models}
Recent work \citep{Lou2020} attempting to use an end-to-end system in place of a two-stage ASR and disfluency detection model found that the end-to-end formulation produced marginally worse results compared to a two-stage approach. The architectures that were used were based on 3 models: A CNN model, a Bahdanau attention LSTM and a Transformer model. Out of the three, the end-to-end Transformer model performed best but failed to match the performance of the two-stage system. At the time of that evaluation, however, large-scale pretraining objectives for audio data were not common, forcing the audio model to learn meaningful feature representations entirely from scratch during training. We explore results on disfluency removal using pretrained models for effective audio representations, which were not evaluated in \citet{Lou2020}.

\section{Data \& Methodology}
\subsection{Dataset}
The dataset used for fine-tuning and evaluation is Switchboard: a collection of 2,400 two-sided telephone conversations between 543 (302 Male, 241 Female) paid volunteers in the United States prompted by a set of 70 discussion topics. The data was collected by Texas Instruments with funding from DARPA \cite{switchboard}. A subset of these conversations was then annotated for syntactic structure and disfluencies by SRI International as part of the Penn Treebank project \cite{Shriberg1996DISFLUENCIESIS} according to the methods laid out in \citet{Shriberg94preliminariesto}.
We will be using these annotations to define our targets for fine-tuning and evaluation. We also make use of transcription and word alignment corrections from \citet{switchboard_release_2}.

\subsection{Dataset Preprocessing}
Annotations for Switchboard, \cite{Shriberg1996DISFLUENCIESIS} are provided in Penn-Treebank format by \citet{zayats} in an XML format containing text tokens, word-aligned time stamps for each conversation in the set and additional metadata for edit, interruption and repair points. 

We make use of code from \citet{nihal-parser-code} and \citet{jonbean-parser-code} to parse the XML files containing the Penn Treebank annotations, which we modify to suit our needs. We produce fluent transcripts by removing sections of text between edit and interruption points and, when applicable, keeping corresponding sections marked as repair. We also remove 'Uh's, 'Um's and tokens ending in "-" which designate words that were cut off before completion. Lastly, we use timestamp annotations to extract audio segments corresponding to each utterance, which are fed to our transcription model and trained using filtered fluent text as target labels.

\subsection{Modeling}

To embed our raw audio file inputs with meaningful representations, we use \citeauthor{HuBERT}'s Hidden-Unit BERT (HuBERT), in particular the version distributed by the HuggingFace API.
The HuBERT model is well suited for the task we are trying to accomplish: it is a self-supervised model that has been demonstrated to produce meaningful audio representations for downstream tasks.

To that end, we use a Conformer architecture \cite{Conformer} pretrained using the offline clustering task defined by HuBERT \cite{HuBERT} to selectively ignore disfluencies using a fine-tuning approach. We leverage the pretrained weights of Facebook's Hubert-large-ls960-ft model, which consists of a HuBERT base with a CTC decoding head fine-tuned on LibriSpeech \cite{panayotov2015librispeech}. Per the original HuBERT paper, we also freeze the weights of the convolution feature encoder during fine-tuning. We filter our audio samples to keep only audio samples 3-15 seconds long and remove special characters from our target text labels, retaining only the text and apostrophes. Additionally, we upsample the 8khz voice recordings from Switchboard to the 16kHz HuBERT expects. 

We also evaluate a two-stage model following the work of \citet{small-bert} by fine tuning a HuBERT ASR model on unfiltered Switchboard transcripts, training a per-token classifier using the Switchboard disfluency annotations and evaluating word and character error rates using disfluency filtered transcripts.

\section{Results}
In our experiments using the Switchboard test set, our end-to-end disfluency model slightly outperforms a two-stage model, achieving a WER of 12.2\% and a CER of 7.3\%, against the two stage model's WER of 13.1\% and CER of 7.6\%.

Additionally, we found that the choice of pretraining objective substantially affects the fine-tuning performance for the disfluency removal task. A pretrained Wav2Vec2 achieves test set performance on WER of 23.9\% and 13.3\%, respectively. This is a significant difference in performance between the two Conformer models. The Wav2Vec2 pretraining objective utilizes a contrastive MLM loss over the full, high dimensional output layer weights; In contrast, HuBERT's auxiliary cluster-prediction pretraining task appears to learn more stable and flexible representations. The HuBERT model significantly outperformed the Wav2Vec2 model on our fine-tuning task, despite the two being competitive on the standard, unmodified ASR task.

%




\section{Ethical Considerations}
\subsection{Dataset Composition}
The Switchboard dataset contains conversations from a limited selection of speakers, all living in the United States, which, according to the original publication, covers "every major dialect of American English". Utterances are annotated with their associated dialects as one of "SOUTHERN", "WESTERN", "NORTHERN", "NEW", "NYC", "MIXED" or "UNK". The limited variety of speech signatures present in the data presents a risk of poor performance on those not appearing in Switchboard.

In addition to the risk of faulty transcriptions for dialects and varieties of speech not present in the pretraining and fine-tuning sets, this model is optimized to reject segments of speech it has determined are "disfluent". In comparison to models that produce full transcripts, incorrect or misinterpreted as they may be, there is the risk that this model not only misinterprets incoming speech but rejects it altogether, distorted or not. For systems that utilize voice as their primary interaction method, this behavior could render them completely inaccessible to speech varieties not covered by the data, and restrict inputs available to downstream systems that utilize its transcriptions of "fluent" speech, as it's determined by our model.

On the other hand, this model also has the potential to make voice systems accessible to users with speech irregularities whose spoken inputs were previously uninterpretable by current systems. The relative risks and benefits of deploying such a system need to be weighed and evaluated before relying on a model that imposes strong priors on the speech they consume.

\subsection{Data Collection and Use}
The data used in Switchboard was collected as part of an effort by DARPA to develop speech recognition, translation and knowledge distillation technologies for their Global Autonomous Language Exploitation (GALE) program. Some of the technologies developed by this program were eventually leveraged in systems to assist American soldiers stationed abroad in communicating with local populations (\citet{gale}; \citet{maeda}). Outcomes of this program include the \textit{IraqComm} system, which was developed and deployed during the Iraq War and used to facilitate two-way conversations between the US military and the local Iraqi population \cite{speech_translation_research}. The production of speech datasets and transcription models are directly tied to the strategic interests of the funding organizations that enable their creation and are inseparable from the downstream systems they ultimately serve. This is true for systems that cover languages with strong representation in current methods, as well as systems for learning representations for underrepresented languages, which can be utilized against the interests of the underrepresented communities they are ostensibly meant to serve.

\section{Conclusion and Next Steps}
In this work, we evaluated the feasibility of training end-to-end ASR and disfluency removal models in light of recent developments in large scale acoustic model pretraining. We showed that Conformer models fine tuned from weights learned during masked audio pretraining can achieve performance on par or better than a two stage approach fine tuned on similar data. We also showed the effect that the choice of pretraining objective can have on the ability of an acoustic model to adapt to a new task.

We note several limitations of this work and potential directions for subsequent work. Firstly, in conducting evaluations on the same data distribution that is used for fine tuning we risk of overstating "in-the-wild" performance of our trained model. This work does not evaluate out-of-distribution performance of our model, and we expect that applications of our model to data distributions not seen during training could result in degenerate performance.

On a more optimistic note, we note that the data used to learn HuBERT's pretrained weights came from a large dataset of non-spontaneous speech; LibriSpeech consists of recordings of audiobooks, which often contain dialogue taking the form of spontaneous speech, but are nonetheless distinct from organic speech. Follow up work could explore the effect that pretraining on a dataset consisting of spontaneous speech, like Mozilla's Common Voice \cite{commonvoice}, could have on a disfluency model.

Finally, a more thorough investigation of the performance of our model between fluent and disfluent sections of speech could bring more nuance to the analysis of the disfluency removal capacities of all of the models evaluated here, which risk being obscured by general ASR metrics like WER and CER. The inclusion of \citet{Lou2020}'s Fluent Error Rate and Disfluent Error rate was planned for this work, but was ultimately omitted due to time constraints. We leave this analysis for later work.

\section{Collaboration Statement}
All team members were active participants in the group's responsibilities to brainstorm project ideas and accumulate related literature. For our project, there were two main coding tasks: modeling and data preprocessing. David was responsible for the contributions to model development, training and evaluation, while Saksham developed the parsing preprocessing library used to transform Penn Treebank formatted data into a representation that could be ingested by our model. He was also responsible for model diagrams. Giulio was responsible for communications with Angelica Chen to establish our baseline two-stage approach, conducting background research, and contributing to the write-up. Siddhartha primarily worked on our two-stage ASR to disfluency detection baseline. He also worked on additional model research.

\bibliographystyle{acl_natbib}
\bibliography{proposal_citations}

\end{document}